\pdfoutput=1
\documentclass[doublecol,final]{epl2}
\usepackage[utf8]{inputenc}
\usepackage[T1]{fontenc}

\usepackage{amsmath}
\usepackage{bm}
\def\BZ{\ensuremath{\text{BZ}}}
\def\dd{\ensuremath{\mathrm{d}}}
\def\ee{\ensuremath{\mathrm{e}}}
\def\ii{\ensuremath{\mathrm{i}}}
\def\un{\ensuremath{\text{I}}}
\def\deux{\ensuremath{\text{II}}}
\usepackage{graphicx}
\usepackage[x11names,svgnames]{xcolor}
\usepackage{hyperref}

\hypersetup{colorlinks=true,citecolor=black,linkcolor=black,urlcolor=black}

\makeatletter
\def\epl@stylemark{}
\def\epl@headrule{}
\def\epl@stretchpretitle{-16ex}
\def\epl@stretchpostrule{-5ex}
\let\old@maketitle\@maketitle
\def\@maketitle{
\def\hrule##1##2{}
\old@maketitle
}

\makeatother

\begin{document}
\title{Parallel Transport and Band Theory in Crystals}
\author{Michel Fruchart \and David Carpentier \and Krzysztof Gaw\k{e}dzki}
\institute{Laboratoire de Physique, \'Ecole Normale Sup\'erieure de Lyon, 
47 all\'ee d'Italie, 69007 Lyon, France}
\date{\today}

\abstract{
We show that different conventions for Bloch Hamiltonians on non-Bravais 
lattices correspond to different natural definitions of parallel transport 
of Bloch eigenstates. Generically the Berry curvatures associated with 
these parallel transports differ, while physical quantities are naturally 
related to a canonical choice of the parallel transport. 
}

\maketitle
An increasing effort has been recently devoted to the characterization 
of geometrical and topological properties of electronic bands in crystals. 
Several physical concepts, including the semi-classical evolution and 
the polarizability, were related through the geometrical notions of Berry 
connection, curvature, and phase to the parallel transport within electronic
bands \cite{Berry1984,SundaramNiu1999,KingSmithVanderbilt1993,XiaoChangNiu2010}. 
It was recently proposed to experimentally access such Berry properties 
in  cold atoms lattices \cite{PriceCooper2012,Atala2012}. Blount already noted 
\cite{Blount1962} that a quantity analogous to an electromagnetic vector 
potential characterizes isolated bands in a crystal. With the advent of 
the integer quantum Hall effect, the first Chern class of 
electronic bands was described via a connection defining parallel transport of 
eigenstates over the Brillouin zone \cite{ThoulessKohmotoNightingaleNijs1982}. 
Independently, Berry \cite{Berry1984} introduced a similar notion 
of parallel transport to describe the adiabatic evolution of eigenstates 
of a time dependent Hamiltonian. Both notions of parallel transport were 
related soon after \cite{Simon1983}, and the properties of parallel 
transport of eigenstates in crystals are now associated with a Berry 
connection, although they are not necessarily related to a Hamiltonian 
adiabatic evolution.

Electronic eigenstates in a crystal can be deduced from a Bloch 
Hamiltonian depending on a quasi-momentum in the Brillouin zone 
\cite{Blount1962}. However, as noted by Zak, this Hamiltonian 
does not have a unique matrix form for a given model on a crystal 
\cite{Zak1989a}.
Several conventions exist
for bases of Bloch states over the Brillouin zone, as recently discussed 
in the context of graphene-like systems in
\cite{FuchsPiechonGoerbigMontambaux2010,BenaMontambaux2009}. 
Such different conventions define different ways of transporting
eigenstates parallel to the bases.
The different parallel transports do not necessarily define 
unique ``Berry geometrical properties'' when restricted to 
a (few) band(s). It is the purpose of this paper to show that different 
``Berry curvatures'' are obtained out of commonly used Bloch conventions for 
models defined on non-Bravais lattices such as graphene. Moreover, we  
point to a canonical Bloch convention that allows to define parallel transport 
in an (almost) unambiguous way. When projected on a (few) band(s), the latter 
defines the physically relevant ``Berry connection'' characterizing 
measurable properties of the crystal. The above situation differs 
from the Hamiltonian adiabatic evolution initially considered 
by Berry where the parallel transport in the whole Hilbert space is 
unambiguously defined from the start. 
   
\begin{figure*}[thb]
	\centering
	\onefigure[width=16cm]{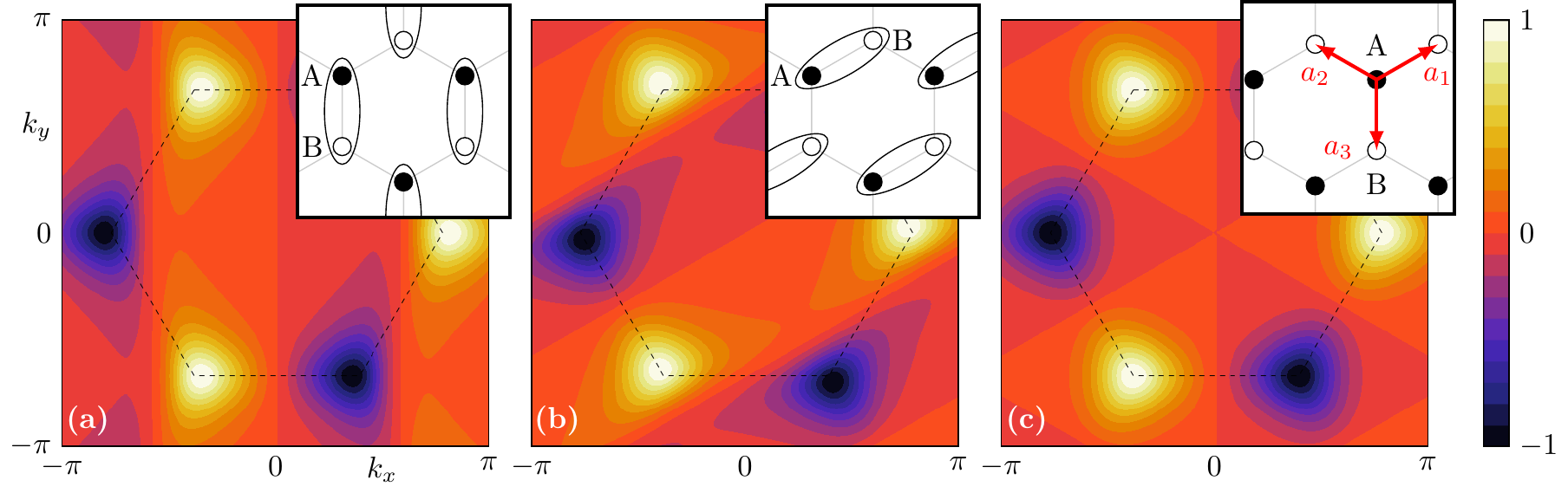}
	\caption{The (normalized) Berry curvature $F_k$ of the valence band 
in a gapped graphene model \eqref{eq:ExampleTBHamiltonian} with $v/t=1$ is 
plotted on the Brillouin zone (dashed hexagone) for conventions I (a and b) 
and II (c). The corresponding
choices of unit cells for convention I are shown in the insets. Vectors
$a_i$ connecting nearest neighbors in sublattices A and B of the graphene
lattice are shown in inset in panel (c). In all three cases, curvature $F_k$ is
concentrated around the Dirac points of graphene. It depends strongly on
the unit cell for convention I and is uniquely defined and respects the
symmetries of the crystal for convention II.}
	\label{fig:BerryColor}
\end{figure*}

Let us begin by considering a crystal $\mathcal C$, a discrete subset of a 
$d$-dimensional Euclidean space ${E}^d$ (the locations of atoms) acted upon 
by a Bravais lattice $\Gamma\subset{\bm R}^d$ of discrete translations, 
$\Gamma\simeq{\bm Z}^d$. For non-Bravais crystals,
several classes of translationally equivalent points of 
$\mathcal C$ called sublattices exist:  
the cardinal $N$ of $\mathcal C/\Gamma$ is larger than $1$. For example, 
the hexagonal lattice of graphene possesses $N=2$ sublattices $A$ and $B$
marked, respectively, as full and empty circles in the insets of 
Fig.~\ref{fig:BerryColor}. The Bravais lattice is composed of vectors
connecting sites of the same sublattice. The set 
$\mathcal C/\Gamma$ may be represented by a fundamental domain 
$\mathcal F\subset\mathcal  C$, also called a unit cell,
which has one point in each sublattice. A translation of $\mathcal F$ 
by a Bravais vector is also a possible choice for a unit cell 
but when $N>1$ then there are choices of $\mathcal F$ that are not related 
in this way. Two such choices for graphene are illustrated in the insets of 
Fig.~\ref{fig:BerryColor}a and b.

Along with the translational degrees of freedom, one may consider a finite 
number of internal degrees of freedom of atoms, e.g.\ the spin of electrons, 
represented as a finite-dimensional complex vector space $V$ equipped with 
a scalar product $\left\langle \cdot,\cdot\right\rangle_{V}$. The 
Hilbert space of crystalline states is then the space $\bm{H}=\ell^{2}
(\mathcal C,V)$ of $V$-valued square-summable functions on 
$\mathcal C$ with the scalar product $\langle \psi|\chi\rangle
=\sum_{x\in\mathcal C}\langle\psi(x)|\chi(x)\rangle_V$. 
The translation of states $\psi\in\bm{H}$ by a vector $\gamma$ of the
Bravais lattice is defined  by the unitary 
operator $T_{\gamma}$ such that $T_\gamma\psi(x)=\psi(x-\gamma)$ for $x\in
\mathcal C$. Below, we shall assume for simplicity that
$V=\bm{C}$ but all considerations generalize in a straightforward
way to the case with internal degrees of freedom.   

Operators $T_\gamma$ define a representation of the translation group 
$\Gamma$ in $\bm{H}$ which may be decomposed into 
irreducible $1$-dimensional components. Irreducible representations 
of $\Gamma$ are its characters $\gamma \mapsto \ee^{\ii k\cdot\gamma}$, where 
$k \in \bm{R}^{d}$. As $\ee^{\ii k\cdot\gamma}=\ee^{\ii(k+G)\cdot\gamma}$ 
for $G$ in the reciprocal lattice $\Gamma^\star$ (composed of vectors $G$ 
with $G\cdot\gamma\in2\pi{\bm Z}$), the characters of $\Gamma$ form 
a $d$-dimensional Brillouin torus $\BZ=\bm{R}^{d}/\Gamma^\star$. The 
decomposition of $\bm{H}$ into irreducible components is realized 
by the Fourier transform $\psi\mapsto\widehat\psi$, where 
\begin{equation}
\widehat\psi_k(x)=\sum_{\gamma\in\Gamma}\ee^{-\ii k\cdot\gamma}
\psi(x-\gamma).
\label{FT}
\end{equation}
Note that $\widehat\psi_k=\widehat\psi_{k+G}$ and 
\begin{equation}
\widehat\psi_k(x-\gamma)=\ee^{\ii k\cdot\gamma}\widehat\psi_k(x)
\label{BF}
\end{equation}
so that $\widehat{T_\gamma\psi}_k(x)=
\ee^{\ii k\cdot\gamma}\widehat\psi_k(x)$.
Property \eqref{BF} defines the Bloch functions on 
$\mathcal C$ with quasi-momentum $k$. For fixed $k$, 
such functions $\varphi_k(x)$ form a finite-dimensional space
$\mathcal{H}_k=\mathcal{H}_{k+G}$ that may 
be equipped with the scalar 
product $\langle\varphi_k|\chi_k\rangle_k=\sum_{\mathcal C/\Gamma}
\overline{\varphi_k(x)}\chi_k(x)$.  
As a Bloch function is determined by its values on a unit cell
$\mathcal F\subset\mathcal{C}$, $\mathrm{dim}(\mathcal{H}_k)=N$. 
Note that spaces $\mathcal{H}_k$ are defined canonically for each $k$,
without any further choices.

Geometrically, the collection of vector spaces $\mathcal{H}_k$ 
forms a complex $N$-dimensional vector bundle 
$\mathcal{H}$ over 
the Brillouin torus $\BZ$. We shall call $\mathcal{H}$ the Bloch 
bundle. Spaces $\mathcal{H}_k$ are the fibers of $\mathcal{H}$ 
and their scalar product equips $\mathcal{H}$
with a Hermitian structure. 
Sections of $\mathcal{H}$ are maps $k\mapsto\varphi_k\in\mathcal{H}_k$.
They are
smooth if functions $k\mapsto\varphi_k(x)$ are smooth for all $x$. 
The Fourier transform \eqref{FT} is an isomorphism between $\bm{H}$ and 
the space of square-integrable sections $k\mapsto\widehat{\psi}_k$ of 
the Bloch bundle $\mathcal{H}$ defined over $\BZ$. It preserves the norm, 
as stated by the Plancherel formula. Its inverse is given by the
normalized integral of $\widehat{\psi}_k(x)$ over the Brillouin torus.

To compare vectors in different fibers along a curve in the base space of a 
bundle one needs a parallel transport. Such a transport is usually not given
{\it a priori}: it requires a prescription, provided by a so-called connection.
Equivalently, a covariant derivative of sections encodes infinitesimal parallel
transport along coordinate axes in the base space. 
Connections always exist but are non-unique. 
A flat connection corresponds to the situation where the parallel transport
of a state along a closed loop always ends in the same state, except for 
loops which cannot be continuously deformed to a point. 
Conversely, for non-flat connections the initial and the parallel transported
states along a contractible closed loop may differ.
Only special bundles carry flat connections. The Bloch bundle does
because it may be trivialized (but not in a canonical way). 
A trivialization of $\mathcal{H}$ is a family of smooth sections
$k\mapsto e^i_k$, $i=1,\dots,N$, defined over $\BZ$ 
(i.e.\,with $e^i_k=e^i_{k+G}$), which for each $k$ form an orthonormal 
basis (a {\em frame}) of $\mathcal{H}_k$.
An example of a trivialization of $\mathcal{H}$ is provided 
by the Fourier transforms of functions $\delta_{x,x_i}$ concentrated 
at points $x_i$, $i=1,\dots,N$, of a fixed unit cell 
$\mathcal F\subset\mathcal C$. We shall denote by $e^{\un i}_k$ 
the corresponding vectors in $\mathcal{H}_k$. Bloch functions decompose as 
$\varphi_k=\sum_i\varphi_k(x_i)e^{\un i}_k$. 
The trivialization of $\mathcal{H}$ 
defined this way depends on the choice of $\mathcal F$.
If $\mathcal F'$ is another unit cell then  
$x'_i=x_i+\gamma_i$ with $\gamma_i\in\Gamma$ for an appropriate 
numbering of its points so that $\delta_{x,x'_i}=T_{\gamma_i}\delta_{x,x_i}$ 
and, consequently, $e'^{\un i}_k=\ee^{\ii k\cdot\gamma_i}e^{\un i}_k$. 

Each trivialization permits to identify the Bloch bundle 
$\mathcal{H}$ with the trivial bundle $\BZ\times\bm{C}^N$ 
and to equip $\mathcal{H}$ with a flat 
connection that we identify with the covariant derivative
$\nabla=\sum \dd{}k_\mu \nabla_\mu$
acting on smooth sections $k\mapsto\varphi_k$ of 
$\mathcal{H}$ by the formula 
\begin{equation}
\langle e^i_k|\nabla\varphi_k\rangle_k=
\dd{}\langle e^i_k|\varphi_k
\rangle_k
\label{conn}
\end{equation}
where $\dd=\sum \dd{}k_\mu\partial_\mu$ is the exterior derivative 
of functions on $\BZ$. Flatness of the connection 
means that $\nabla^2=0$. It follows from the property $\dd^2=0$ of the
exterior derivative. We shall denote by $\nabla^{\un}$ the flat connection 
associated to the trivialization $k\mapsto e^{\un i}_k$ of 
$\mathcal{H}$ defined above. $\nabla^{\un}$ depends on the 
choice of unit cell $\mathcal F$ but is independent of the
numbering of its points. For another unit cell $\mathcal F'$
related to $\mathcal{F}$ as discussed above, we have 
${\nabla'^{\un}}=\nabla^{\un}-\ii\sum_i|e_k^{\un i}\rangle\langle e_k^{\un i}|
\,\dd{}k\cdot\gamma_i$. The difference is a differential 1-form with values in
linear transformations of the fibers of $\mathcal{H}$.

There exists a more canonical and physically more relevant way to equip 
the Bloch bundle with a flat connection. Upon a choice of the origin 
$x_0$ in the Euclidean space $E^d$, we may identify Bloch functions with 
$\Gamma$-periodic functions on $\mathcal C$ by writing for any 
$\varphi_k \in \mathcal{H}_k$,
\begin{equation}
\varphi_k(x)=\ee^{-\ii k\cdot(x-x_0)}u_k(x).
\label{PF}
\end{equation}
Clearly $u_k(x+\gamma)=u_k(x)$ and $u_{k+G}(x)
=\ee^{\ii G\cdot(x-x_0)}u_k(x)$ for $\gamma\in\Gamma$ and $G\in\Gamma^\star$.
This allows to identify the Bloch bundle 
with the quotient of the trivial bundle $\bm{R}^d\times \ell^2(\mathcal 
C/\Gamma)$ over $\bm{R}^d$, with the fiber composed of $\Gamma$-periodic
functions $u(x)$ on $\mathcal{C}$, by the action 
\begin{equation}
(k,u(x))\longmapsto(k+G,\,\ee^{\ii G\cdot(x-x_0)}u(x))
\end{equation}
of the reciprocal lattice $\Gamma^\star$. Note that an element 
$G\in\Gamma^\star$ 
acts on $u(x)$ by multiplying it by the $\Gamma$-periodic function 
$\ee^{\ii G\cdot(x-x_0)}$, preserving the $\ell^2(\mathcal C/\Gamma)$ scalar 
product. The trivial bundle $\bm{R}^d\times \ell^2(\mathcal C/\Gamma)$ has 
a natural flat connection given by the exterior derivative of its sections. 
As the functions $\ee^{\ii G\cdot(x-x_0)}$ do not depend on $k$, this 
connection 
commutes with the action of $\Gamma^\star$ and, consequently, it induces 
a connection $\nabla^{\deux}$ on the Bloch bundle $\mathcal{H}$. The induced
connection is still flat: $(\nabla^{\deux})^2=0$, but it has 
a non-trivial holonomy along the non-contractible loops of the Brillouin 
torus $\BZ$. This holonomy may be identified via relation \eqref{PF} with 
the above action of $\Gamma^\star$ on $\ell^2(\mathcal C/\Gamma)$. 
Physical importance of connection $\nabla^\deux$ resides in the fact that 
$-\ii\nabla^\deux_\mu$ corresponds under the Fourier transform to the position 
operator multiplying wave functions $\psi(x)$ by $(x-x_0)_\mu$. 
We can still use  eq.~\eqref{conn} to define $\nabla^{\deux}$ but now with  
Bloch functions $e^i_k(x)=\ee^{-\ii k\cdot(x-x_0)}u^i(x)$, where $u^i$, 
$i=1,\dots,N$, form any orthonormal basis of the space 
$\ell^2(\mathcal C/\Gamma)$ of $\Gamma$-periodic functions on $\mathcal C$
(note that $e^i_{k+G}\not=e^i_k$ in this case).
For the particular choice with $u^i(x)=\sum_{\gamma\in\Gamma}
T_\gamma\delta_{x,x_i}$, where $x_i$ are the points of a unit cell 
$\mathcal F$, we shall denote the corresponding Bloch functions
by $e^{\deux i}_k(x)$. Although connection $\nabla^{\deux}$ does 
not depend on the choice of $\mathcal F$, it does depend on the choice 
of the origin $x_0$ of the Euclidean space, but in a very simple way. If 
we choose another origin $x'_0$ then $e'^{\deux i}_k=\ee^{\ii k\cdot(x'_0-x_0)}
e^{\deux i}$ so that 
$\nabla'^{\deux}=\nabla^{\deux}-\ii\,\dd{}k\cdot(x_0'-x_0)$, 
i.e.\ the two connections differ by a closed scalar $1$-form. 
To compare connections $\nabla^\un$ and $\nabla^\deux$, one notes that
$e^{\deux i}_k=\ee^{-\ii k\cdot(x_i-x_0)}e^{\un i}_k$ for the same choice of 
$\mathcal F$.
Hence $\nabla^\deux=\nabla^\un-\ii\sum_i|e^{\un i}_k\rangle\langle e^{\un i}_k|
\,\dd{}k\cdot(x_i-x_0)$.
  
Often we restrict our attention to electronic states in a $\Gamma$-invariant 
subspace of $\bm{H}$, such as valence bands in an insulator. This amounts 
to considering a subbundle $\mathcal{E}$ of the Bloch bundle, 
{\it i.e.} a collection of $M$-dimensional vector 
subspaces $\mathcal{E}_k\subset\mathcal{H}_k$ smoothly dependent on $k$ 
and such that $\mathcal{E}_{k+G}=\mathcal{E}_{k}$. 
Any connection $\nabla$ on the bundle $\mathcal{H}$ projects 
to a connection ${}^\mathcal{E\hspace{-0.03cm}}\nabla$ on $\mathcal{E}$ 
for which the covariant derivative of sections $k\mapsto\varphi_k\in
\mathcal{E}_k$ is given by 
\begin{equation}
{}^\mathcal{E\hspace{-0.03cm}}\nabla\varphi_k
=\mathcal{P}_k\nabla\varphi_k
\end{equation}
where $\mathcal{P}_k$ is the orthogonal projector from 
$\mathcal{H}_k$ to $\mathcal{E}_k$.
In general, flat connections project to connections with curvature.  
In particular, if $\nabla$ is a flat connection obtained 
by formula \eqref{conn} then
$\langle e^i_k|{}^\mathcal{E\hspace{-0.03cm}}
\nabla\varphi_k\rangle_k=\sum_jP^{ij}_k
\dd{}\langle e^j_k|\varphi_k\rangle_k$ where $P^{ij}_k=\langle e^i_k|
\mathcal{P}_k|e^j_k\rangle_k$. In this case
\begin{equation}
\langle e^i_k|{}^\mathcal{E\hspace{-0.03cm}}
\nabla^2\varphi_k\rangle_k=\sum_j F^{ij}_k
\langle e^j_k|\varphi_k\rangle_k
\end{equation}
where $\,F^{ij}=\sum_{mn}P^{im}\dd{}P^{mn}\hspace{-0.05cm}\wedge\dd{}P^{nj}$ 
is the matrix curvature 
2-form and $F=\sum_i F^{ii}={\rm tr}\,(P\dd{}P\hspace{-0.05cm}\wedge\dd{}P)$ 
is its scalar version.

The integrals divided by $-2\pi\ii$ of $F$ over $2$-dimensional (subtori 
in) $\BZ$ give the $1^{\rm st}$ Chern number(s) of the vector bundle 
$\mathcal{E}$ that are independent of the
choice of connection. We may work, in particular, with the connections
${}^\mathcal{E\hspace{-0.03cm}}\nabla^{\un}$ or ${}^\mathcal{E\hspace{-0.03cm}}
\nabla^{\deux}$ obtained by projection
of $\nabla^\un$ and $\nabla^\deux$ to $\mathcal{E}$. 
Scalar curvature $F^\un$ corresponding to connection 
${}^\mathcal{E\hspace{-0.03cm}}\nabla^{\un}$ 
depends in general on the unit cell 
$\mathcal{F}\subset\mathcal{C}$ whereas scalar curvature $F^\deux$ 
corresponding to connection ${}^\mathcal{E\hspace{-0.03cm}}
\nabla^{\deux}$ does not depend on $x_0$ and is canonically defined 
for each subbundle $\mathcal{E}\subset\mathcal{H}$. To see this, note 
that the difference between 
different frames of sections $k\mapsto e^i_k$ and $k\mapsto e'^i_k$ involved 
in the definitions \eqref{conn} has the form $e^i_k=\sum_jU^{ji}_ke'^j_k$ 
for unitary matrices $U_k$. The corresponding projectors $P'_k$ 
and $P_k$ given by matrices $P'^{ij}_k$ and $P^{ij}_k$ are related by 
the equality $P'_k=U_kP_kU_k^{-1}$ leading to the 
relation 
\begin{equation}
F'=F+{\rm tr}\,(\dd{}P\wedge U^{-1}\dd{}U - P\,U^{-1}\dd{}U\wedge U^{-1}\dd{}U).
\end{equation}
In general, operators $U$ do not commute with $P$ and we obtain 
different scalar curvatures. An exception is the relation between 
the sections $k\mapsto e^{\deux i}_k$ corresponding 
to different $x_0$ where operators $U$ are scalar and $P'=P$ 
resulting in the same scalar curvatures. 

We shall use the above geometric setup for the case of
the subbundle$\mathcal{E} \subset\mathcal{H}$ of the valence-band states
of an insulator. To this end, let us consider a tight-binding Hamiltonian
\begin{equation}
	H=\sum_{x,y\in\mathcal{C}}h_{x,y}|x\rangle\langle y|
\end{equation}
in the Hilbert space $\bm{H}$, where $|y\rangle$ denotes the state 
with localized wave function $x \mapsto \delta_{x,y}$. We shall assume that 
$h_{x,y}=\overline{h_{y,x}}$ 
vanishes for $|x-y|$ outside some fixed range. If 
$h_{x,y}=h_{x+\gamma,y+\gamma}$ for $\gamma\in\Gamma$ then $H$ commutes 
with Bravais lattice translations $T_\gamma$ and maps Bloch functions 
into Bloch functions, defining Bloch Hamiltonians $H_k=H_{k+G}$ acting in 
the finite-dimensional spaces $\mathcal{H}_k$.
Under the Fourier transform \eqref{FT},
\begin{equation}
\widehat{H\psi}_k=H_k\widehat{\psi}_k.   
\end{equation}
Given a frame of sections $k\mapsto e^i_k$ of the Bloch bundle 
$\mathcal{H}$, Hamiltonians $H_k$ may be represented by 
$N\times N$ Bloch matrices $\langle e^i_k|H_k|e^j_k\rangle_k
\equiv H_k^{ij}=\overline{H_k^{ji}}$ related by unitary transformations 
for different choices of the frame but describing the same physics. 
In particular, we may obtain Bloch matrices $H^{\un}_k=H^\un_{k+G}$ 
corresponding to the choice of frames 
$e^{\un i}_k$ or $H^{\deux}_k\not=H^{\deux}_{k+G}$ corresponding 
to the frames $e^{\deux i}_k$, a situation considered 
in \cite{Zak1989b}. These matrices are the usual standard forms of Bloch 
Hamiltonians.
In the context of graphene, they correspond
to the two conventions 
discussed 
in \cite{BenaMontambaux2009}.
The spectrum $E_{k1}\leq\dots\leq E_{kN}$ of the Bloch matrices is independent 
of the frame and coincides with the spectrum of operators $H_k$. 
For insulators, the Fermi energy $\epsilon_F$ has a value that lies in 
the spectral gap, i.e.\ it differs from all $E_{kn}$.
The subspaces $\mathcal{E}_k\subset\mathcal{H}_k$ corresponding
to $E_{kn}<\epsilon_F$ form the valence-band subbundle $\mathcal{E}$ of
the Bloch bundle $\mathcal{H}$. The space of sections of $\mathcal{E}$ 
is mapped by the inverse Fourier transform to the subspace 
$\bm E\subset\bm H$ of the electronic states with energy $<\epsilon_F$ that
are filled at zero-temperature. The geometrical properties of bundle 
$\mathcal{E}$ have a bearing on the low 
temperature physics of the insulator. The previous discussion applies 
directly to the valence-band subbundle. In particular, 
we may equip $\mathcal{E}$ with different connections 
${}^\mathcal{E\hspace{-0.03cm}}\nabla^{\un}$ 
or ${}^\mathcal{E\hspace{-0.03cm}}\nabla^{\deux}$ whose scalar curvatures 
differ but give rise to the same Chern numbers. 

We can express the projected connections
${}^\mathcal{E\hspace{-0.03cm}}\nabla$ in terms of a local frame 
$k\mapsto\varphi^a_k=\sum_i\varphi^a_ie^i_k$, $a=1,\dots,M$, of $\mathcal{E}$,
{\it e.g.} composed of the eigenstates of Hamiltonians $H_k$ with energies 
$E_{ka}<\epsilon_F$ over regions in $\BZ$ where no such energy levels cross. 
For $\nabla$ given by \eqref{conn}, one has:
\begin{equation}
\langle\varphi^a_k|{}^\mathcal{E\hspace{-0.03cm}}\nabla\varphi^b_k
\rangle_k=
\sum_i\overline{\varphi^a_{ki}}\dd{}\varphi^b_{ki}\equiv A^{ab}_k\label{BC}
\end{equation}
where $A^{ab}=-\overline{A^{ba}}$ is the local connection 1-form, 
and $F={\sum}_a\dd{}A^{aa}$. 
The same relation defines the {\em Berry connection} extracted 
from the change of eigenstates under adiabatic changes of the Hamiltonian.
Although there is usually no underlying physical adiabatic process involved 
in the definition of connections projected on the valence-band subbundle, 
those are often dubbed ``Berry connections''. 
We may just talk of Berry 
connections ${}^\mathcal{E\hspace{-0.03cm}}\nabla^{\un}$ and 
${}^\mathcal{E\hspace{-0.03cm}}\nabla^{\deux}$ on  
$\mathcal{E}$. In particular, the (almost) canonical connection 
${}^\mathcal{E\hspace{-0.03cm}}\nabla^{\deux}$ is related to the position 
operator projected 
on the subspace $\bm{E}=P^{-}\bm{H}$ of states with energy $<\epsilon_F$:
the operator $(x-x_0)_\mu^{-}\equiv P^{-}(x-x_0)_\mu 
P^{-}$ corresponds under 
the Fourier transform to the covariant derivative 
$-\ii{\hspace{0.02cm}}^\mathcal{E\hspace{-0.03cm}}\nabla^{\deux}_\mu$ 
and, for a single 
valence band, the commutator $[(x-x_0)_{\mu}^{-},(x-x_0)_{\nu}^{-}]$ 
measuring the non-commutativity of the projected position operators 
corresponds to the multiplication by the component $F^{\deux}_{\nu\mu}$ of 
the canonical scalar curvature $F^{\deux}$. One may trace back the occurrence 
of Berry connection $\nabla^{\deux}$ in physical properties 
\cite{XiaoChangNiu2010,KingSmithVanderbilt1993} to the above relations.

We now illustrate the different choices of Berry connections on the gapped 
graphene model with an alternate potential on different sublattices.
The model is described by the Hamiltonian
\begin{equation}
H=
t \sum_{\langle x,y\rangle}|x \rangle\langle y| 
+\sum_{x\in\mathcal{C} } v_x|x\rangle\langle x| 
\label{eq:ExampleTBHamiltonian}
\end{equation}
where $\langle x,y\rangle$ run through the nearest neighbor pairs on the
hexagonal crystal $\mathcal{C}$ and
$v_x=+v$ (resp. $-v$) on its sublattice $A$ (resp. $B$). 
In convention $\un$, for the unit cell 
$\mathcal{F}=\{x_A,x_B\}$ shown in Fig.~\ref{fig:BerryColor}a, the Bloch 
matrix Hamiltonians are
\begin{equation}
	H^{\un}_k = \, \begin{pmatrix}
		+v & \overline{g^{\un}_k} \\
		g^{\un}_k & -v 
	\end{pmatrix}
	\label{eq:ExampleBlochHamiltonian}
\end{equation}
with
$g^{\un}_k=t[1+\exp(\ii k\cdot b_1)+\exp(-\ii k\cdot b_2)]$, where 
$b_{i}=\varepsilon_{ijk}(a_{j}-a_{k})$ are the Bravais vectors between 
second nearest neighbors and  $a_i$, $i=1,2,3$, the ones between nearest 
neighbors shown on the inset in Fig.~\ref{fig:BerryColor}c.
A second choice of unit cell $\mathcal{F}'$ shown 
in Fig.~\ref{fig:BerryColor}b leads to  
$g'^{\un}_k=t[1+\exp(\ii k\cdot b_2)+\exp(-\ii k\cdot b_3)]$. 
These Hamiltonians are periodic in $k$ but they explicitly depend on 
the choice of unit cell. 
The Hamiltonians $H^\deux_k$ in canonical convention $\deux$ can 
be deduced from $H^\un_k$ using
the change of basis matrix $U_k=\ee^{\ii k\cdot x_0}\operatorname{diag}
(\ee^{-\ii k\cdot x_A},\ee^{-\ii k\cdot x_B})$ with an arbitrary $x_0$. 
It takes the form \eqref{eq:ExampleBlochHamiltonian} but with $g_k^I$
replaced by $g^{\deux}_k=\ee^{\ii k\cdot a_3}g^{\un}_k=t[\exp(\ii k\cdot a_1)
+\exp(\ii k\cdot a_2)+\exp(\ii k\cdot a_3)]$.
In all cases the spectrum is $E_{k\pm}=\pm(v^2+|g_k|^2)^{1/2}$ and, for
$\epsilon_F=0$, the valence band corresponds to the minus sign. Curvature 
$F$ of the valence-band subbundle has only one component $F_{12}$. It 
is represented in Fig.\ \ref{fig:BerryColor} 
for the three above conventions for the 
gapped graphene with unit $v/t$. The dependence of 
the Berry curvature on the Bloch conventions is clearly 
illustrated. In particular, for convention I, a rotation of the unit 
cell rotates the curvature plot while a translation of the unit cell 
amounts to a $U(1)$ gauge transformation and does not change 
the curvature. 

This ends our discussion of the two main conventions used to define
Berry connections in subbundles of Bloch states, the main subject 
of the present paper. We showed how in one of those conventions the scalar 
curvature depends on additional choices and we identified another, more
physical, convention in which the scalar curvature is unambiguously defined
and relates to the position operator.

\begin{acknowledgments}
\indent\emph{Acknowledgments}. D.C. and M.F. thank B. Dou\c{c}ot, 
J.-N. Fuchs, G. Montambaux and F.\ Pi\'echon for insightful discussions.
This work was supported by a grant from the Agence Nationale de la Recherche 
(ANR Blanc-2010 IsoTop).
\end{acknowledgments}

\bibliographystyle{eplbib}
\bibliography{biblio_ptbtc}

\begin{thebibliography}{10}
\expandafter\ifx\csname url\endcsname\relax\def\url#1{\texttt{#1}}\fi

\bibitem{Berry1984}
\Name{Berry M.~V.} \REVIEW{Proc. Royal Soc. Lond. A. Math. Phys.
  Sci.}{392}{1984}{45}.

\bibitem{SundaramNiu1999}
\Name{Sundaram G. \and Niu Q.} \REVIEW{Phys. Rev. B}{59}{1999}{14915}.

\bibitem{KingSmithVanderbilt1993}
\Name{{King-Smith} R.~D. \and {Vanderbilt} D.} \REVIEW{Phys. Rev.
  B}{47}{1993}{1651}.

\bibitem{XiaoChangNiu2010}
\Name{{Xiao} D., {Chang} M.-C. \and {Niu} Q.} \REVIEW{Rev. Mod.
  Phys.}{82}{2010}{1959}.

\bibitem{PriceCooper2012}
\Name{{Price} H.~M. \and {Cooper} N.~R.} \REVIEW{Phys. Rev.
  A}{85}{2012}{033620}.

\bibitem{Atala2012}
\Name{{Atala} M., {Aidelsburger} M., {Barreiro} J.~T., {Abanin} D., {Kitagawa}
  T., {Demler} E. \and {Bloch} I.} \REVIEW{Nature Phys.}{9}{2012}{795}.

\bibitem{Blount1962}
\Name{Blount E.} \Book{Formalisms of band theory} in proc. of \Book{Solid State
  Physics}, edited by \Name{Seitz F. \and Turnbull D.} Vol.~13 (Academic Press)
  1962 pp. 305--373.

\bibitem{ThoulessKohmotoNightingaleNijs1982}
\Name{Thouless D.~J., Kohmoto M., Nightingale M.~P. \and den Nijs M.}
  \REVIEW{Phys. Rev. Lett.}{49}{1982}{405}.

\bibitem{Simon1983}
\Name{Simon B.} \REVIEW{Phys. Rev. Lett.}{51}{1983}{2167}.

\bibitem{Zak1989a}
\Name{Zak J.} \REVIEW{Phys. Rev. Lett.}{62}{1989}{2747}.

\bibitem{FuchsPiechonGoerbigMontambaux2010}
\Name{{Fuchs} J.~N., {Pi{\'e}chon} F., {Goerbig} M.~O. \and {Montambaux} G.}
  \REVIEW{Eur. Phys. J. B}{77}{2010}{351}.

\bibitem{BenaMontambaux2009}
\Name{Bena C. \and Montambaux G.} \REVIEW{New J. Phys.}{11}{2009}{095003}.

\bibitem{Zak1989b}
\Name{Zak J.} \REVIEW{Europhys. Lett.}{9}{1989}{615}.

\end{thebibliography}
\end{document}